\documentclass[journal]{IEEEtran}

\usepackage{colortbl}
\usepackage[pdftex]{graphicx}

\ifCLASSINFOpdf

\else

\fi

\usepackage[cmex10]{amsmath}

\usepackage{amsmath, amsthm, amssymb, amsfonts, array}
\usepackage{graphicx}
\usepackage{caption}
\usepackage{subcaption}

\begin{document}
\title{Understanding the IoT Connectivity Landscape --\\A Contemporary M2M Radio Technology Roadmap}
%
%
%

\author{Sergey~Andreev$^{\dagger}$, 
				Olga~Galinina, 
				Alexander~Pyattaev, 
        Mikhail~Gerasimenko,\\ 
        Tuomas~Tirronen, 
        Johan~Torsner, 
        Joachim~Sachs, 
        Mischa~Dohler, 
        and~Yevgeni~Koucheryavy
\thanks{S.~Andreev, O.~Galinina, A.~Pyattaev, M.~Gerasimenko, and Y.~Koucheryavy are with the Department of Electronics and Communications Engineering, Tampere University of Technology, FI-33720 Tampere, Finland.}
\thanks{T.~Tirronen, J.~Torsner, and J.~Sachs are with Ericsson, Finland and Sweden.}
\thanks{M.~Dohler is with King's College London, UK; and Worldsensing, UK and Spain.}
\thanks{This work is supported by GETA, TISE, and the Internet of Things program of DIGILE, funded by Tekes. The work of the first author is supported with a Postdoctoral Researcher grant by the Academy of Finland as well as with a Jorma Ollila grant by Nokia Foundation.}
\thanks{$^{\dagger}$S.~Andreev is the contact author: Room TG417, Korkeakoulunkatu 1, 33720, Tampere, Finland (+358 44 329 4200); e-mail: sergey.andreev@tut.fi}
\thanks{\textbf{September 2015}; 00102-Internet of Things, administrated by Milizzo, Joseph}
}

%

\maketitle

\begin{abstract}
This article addresses the market-changing phenomenon of the Internet of Things (IoT), which relies on the underlying paradigm of machine-to-machine (M2M) communications to integrate a plethora of various sensors, actuators, and smart meters across a wide spectrum of businesses. The M2M landscape features today an extreme diversity of available connectivity solutions which $-$ due to the enormous economic promise of the IoT $-$ need to be harmonized across multiple industries. To this end, we comprehensively review the most prominent existing and novel M2M radio technologies, as well as share our first-hand real-world deployment experiences, with the goal to provide a unified insight into enabling M2M architectures, unique technology features, expected performance, and related standardization developments. We pay particular attention to the cellular M2M sector employing 3GPP LTE technology. This work is a systematic recollection of our many recent research, industrial, entrepreneurial, and standardization efforts within the contemporary M2M ecosystem.
\end{abstract}

\section{Introduction and Opportunities}
In the 90s, the word ``Internet'' had the connotation of a physical system of computers networked by means of an Ethernet cable; today, this is forgotten and the Internet is synonymous with the likes of Facebooks, eBays, and LinkedIns. The Internet has thus undergone an enormous transformation from being technology-driven to becoming market-driven. The decoupling of underlying technologies from the services able to run on top of them has been a painful but instrumental shift in unlocking what is now often referred to as the 3rd Industrial Revolution.

\begin{figure*}[!ht]
\centering
\includegraphics[width=2.0\columnwidth]{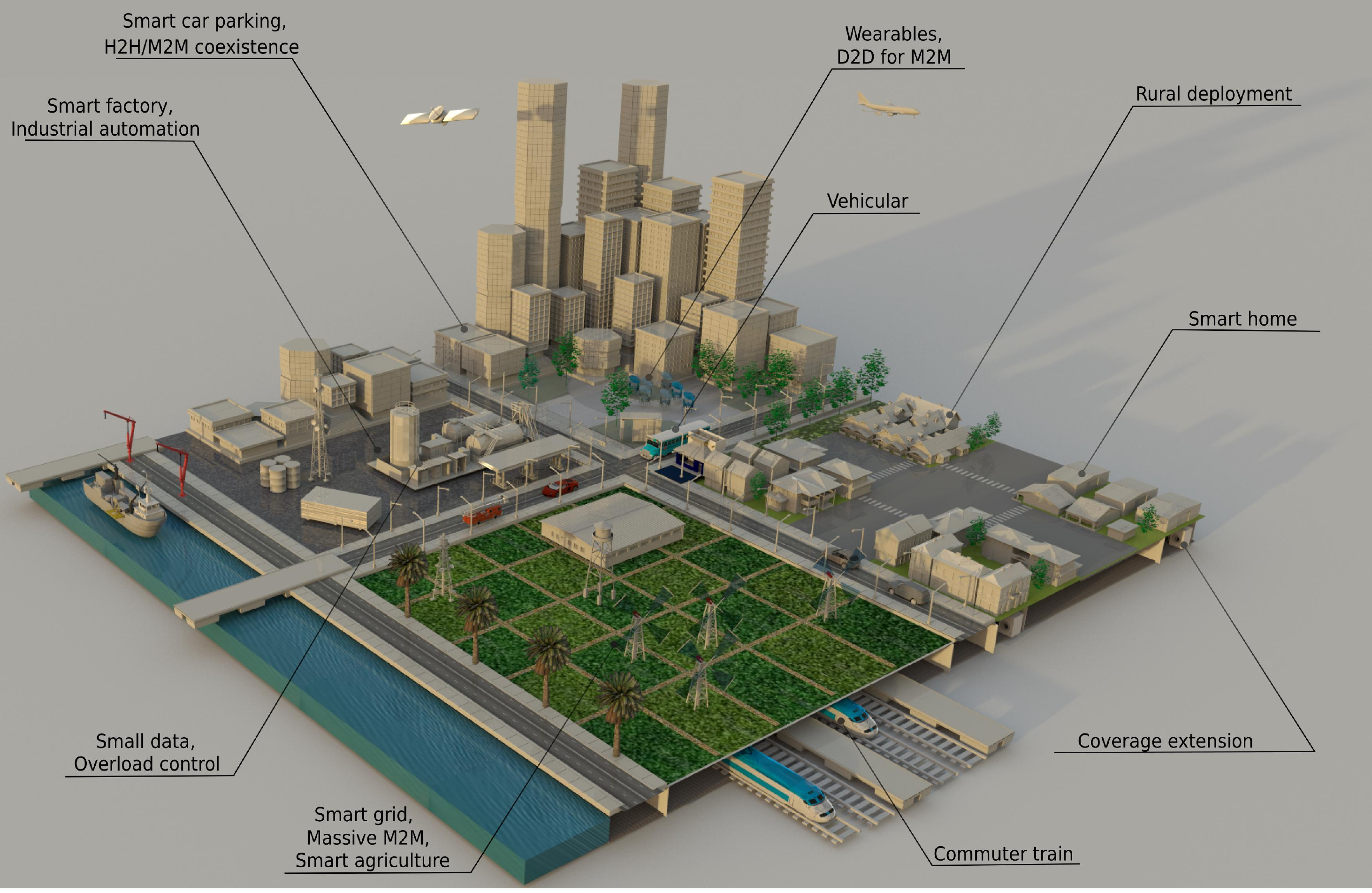}
\caption{A vision of diverse M2M use cases across a variety of industries.}
\label{fig:city}
\end{figure*}

Going beyond the 3rd Industrial Revolution, we are rapidly moving towards a world of ubiquitously connected objects, things, and processes. It is the world of the emerging Internet of Things (IoT), which has the potential to produce a new wave of technological innovation. Indeed, the range of IoT applications is extremely broad, from wearable fitness trackers to connected cars, spanning the industries of utilities, transportation, healthcare, consumer electronics, and many others (see Fig.~\ref{fig:city}). 

However, we are only beginning to witness the true explosive growth of the IoT, with 10 billion M2M devices connected presently and 24 to 50 billion total connections expected within the next 5 years. Over the following decade, we may thus see our everyday furniture, food containers, and even paper documents accessing the Internet. Futurists have also coined a number of new keywords to emphasize the IoT's ongoing transformation, including the Internet of Everything (by Cisco), the Industrial Internet (by General Electric et al.), as well as the Networked Society (by Ericsson).

Today, Machine-to-Machine (M2M) technologies are an integral part of the IoT connectivity ecosystem~\cite{SRA11} and serve as the underlying facilitator for the IoT phenomenon. But they are just a small part; they are the beginning; they are, in a sense, the new (mostly wireless and feature-richer) ``Ethernet cable'' able to connect objects with other objects, with people, and the enormous computing nervous system spanning the globe. Surprisingly, the design efforts related to M2M span back a few decades.

Indeed, driven by industrial needs, early forms of M2M connectivity trace back to supervisory control and data acquisition (SCADA) systems of the 1980s, all being highly isolated and proprietary connectivity islands~\cite{Sto14}. Along the way of its rapid development, the connectivity landscape has embraced legacy Radio Frequency Identification (RFID) technologies (starting in the late 80s), as well as Wireless Sensor Network (WSN) technology (starting in the 90s). Marked by the very attractive application scenarios in both business and consumer markets, the first decade of the 21st century was thus dedicated to the development of standardized low-power M2M solutions, through either industry alliances or standards developing organizations (SDOs).

Notable examples tailored to a range of industry verticals are ISA100.11a, WirelessHART, Z-Wave, and KNX. More generic (horizontal) connectivity technologies were developed within the leading SDOs, i.e. the IEEE, ETSI, 3GPP, and IETF (even though strictly not an SDO). Low-power short-range solutions available today include Bluetooth (promoted by the Bluetooth SIG) and IEEE 802.15.4 (promoted by the Zigbee alliance)~\cite{Sac14}. In subsequent years, the IEEE 802.15.4 physical (PHY) and medium access control (MAC) layers have been complemented by the IP-enabled (networking), as well as the web-enabled IETF stacks. In parallel, capitalizing on the ability to provide global coverage, 3GPP developed cellular-enabled machine-type connectivity modules~\cite{Ast13} tailored to markets with inherent mobility (e.g., car telemetry).

Despite decade-long developments by some of the best engineering teams in the world, none of the above technologies has emerged as a clear market leader. The reasons $-$ a mix of technology shortcomings and business model uncertainties $-$ are rather important and thus discussed subsequently. A key consequence, however, is that the field of the IoT connectivity is now at a turning point with many promising radio technologies emerging as true M2M connectivity contenders: Low-Power WiFi, Low-Power Wide Area (LPWA) networks, and various improvements for cellular M2M systems. 

These afterthought solutions may be significantly more attractive for the prospective IoT deployments from both availability and reliability points of view, and $-$ given their emerging nature $-$ we focus on characterizing these in the remainder of this article. With our hands-on experience in design, standardization, as well as roll-out of these technologies, we share our most essential findings in this work. We believe that these solutions may allow for a decisive transformation of the global M2M industry and thus enable a truly dynamic and sustainable IoT ecosystem at par with the Internet of today.

\section{Smart City IoT $-$ The Awakening Reality Check}

\begin{figure*}[!ht]
\centering
\includegraphics[width=2.0\columnwidth]{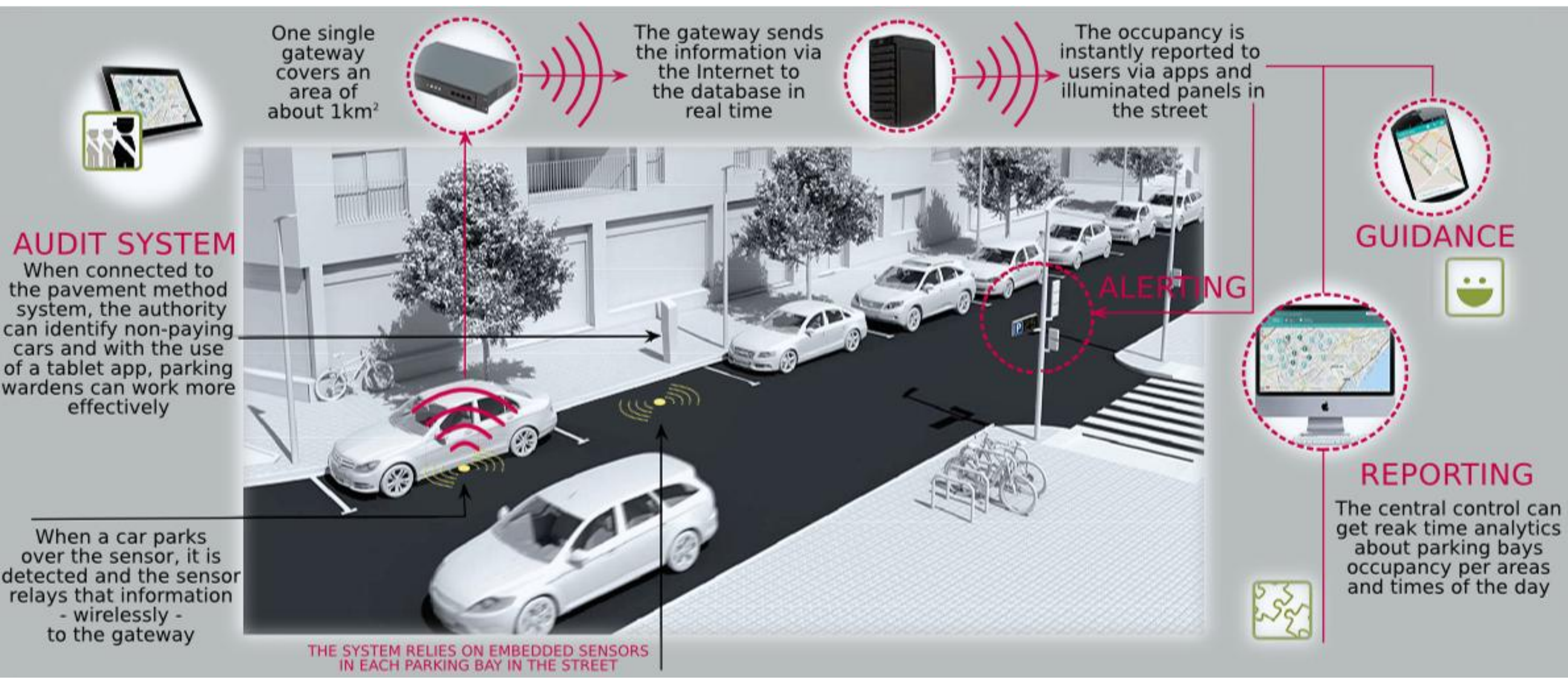}
\caption{Functionality of Worldsensing's real-world IoT deployment of smart parking technologies.}
\label{fig:smartparking}
\end{figure*}

From 2010-2012, we have been using a set of aforementioned technologies in various Smart City deployments around the world. After a decade of theoretical, design, standards, and engineering work, it was thus an opportunity to prove the viability of IoT connectivity solutions. We share our experience in what follows, which forms the rationale for the subsequent sections.

\subsection{Real-World IoT Roll-Outs}
Worldsensing pioneered the concept of Smart Parking which, as shown in Fig.~\ref{fig:smartparking}, involves the placement of sensors in every parking space to detect the presence or absence of cars in real-time~\cite{Moscow}. This information is relayed to the driver who thus avoids circulating in the city in a quest for a parking space. 

Smart parking systems are of interest to cities as they reduce pollution, traffic volumes, and thus traffic jams. Given that congestion currently costs Europe about 1\% of its GDP every year, smart parking is also seen as boosting the economy. Above projections, however, unlock only if some technical key performance indicators (KPIs) are met. Notably, the system should not be in outage for more than 0.1\% per annum which translates to about 9h per year. Furthermore, the parking information needs to be relayed within a few seconds. 

Over the years of 2010-2012, the smart parking systems of various sizes have been trialed in various cities, such as Moscow, Barcelona, Sant Cugat (satellite city to Barcelona), among others. The topology of these early roll-outs included the Zigbee-powered smart parking nodes connected in a multihop mesh network until a repeater. The Zigbee-powered repeaters also networked in a multihop fashion until the gateway. The gateway was connected to the Internet via an Ethernet connection or a cellular 3G modem. There was a repeater every 5-10 parking nodes; and a gateway every 100-150 parking nodes. A trial typically involved at least 100 live parking spaces in a city, thereby giving a realistic picture of the technology's capabilities at a reasonable network scale.

\subsection{Observed IoT Deployment Challenges}

The first challenge was to identify suitable locations to place repeaters and gateways. Given the sheer density of repeaters needed and the high uncertainty of the propagation conditions, this involved a very complex planning. A further uncertainty was around who pays the electricity for repeaters and gateways. To put this into perspective, a city like Moscow with 60K parking spaces would require 6K lamp posts to be equipped with repeaters.

The second challenge was to ensure reliable and robust connectivity no matter what the car parking and interference situation is, so as to ensure the above KPIs to be met. Given the high degree of freedom of the network due to mesh deployment and the high dynamics of the channel due to cars' movement, the system was very often in outage. In some unfavorable testing conditions, the outage reached 10\% or more and would thus have been in straight violation of a service level agreement (SLA).

The third challenge was to ensure that the KPIs in terms of delay were met. Given the multihop nature and the high channel dynamics with frequent outages, delays of minutes often occurred. Again, this would have violated any SLA.

\subsection{Lessons Learned}

Arguably, the biggest mistake of the M2M community in the early days was to believe in the need for low-power technologies, when we actually needed a high-transmission-power low-energy solution. This seems contradictory but remember that power gives you range and energy drains your battery; and energy = power x time. IEEE 802.15.4 systems, however, only offer low power which leads to short transmission range, and thus the need for multihop; this, in turn, yields poor reliability due to the high degree of freedom as we have experienced in the above-described roll-outs. WiFi/3GPP technologies, transmitting at high power (with the advantage of a high communication range), are able to be energy efficient as long as the transmission is done within the shortest time. 

Another lesson we learned is that system reliability matters, and not only link reliability. To be able to rely on a functioning M2M deployment, the underlying end-to-end system must be reliable and available; and not only singular links. In light of these requirements, it is apparent that coverage, support of mobility, and roaming are very poor with Zigbee, at least in large-scale Smart City deployments. Given that Zigbee did and still (mid-2015) enjoys roll-outs, it is far from reaching a critical mass. From a reliability point of view, it is extremely susceptible to interference (particularly in urban environments), has no throughput guarantees, and often produces lengthy system outages. Major companies have finally realized this and stopped producing Zigbee chips, whilst ramping up on low-power WiFi chip ranges.

\section{Emerging M2M Technologies}
Whereas Zigbee-like solutions may still find their market niche with simple and static applications, the large IoT market is $-$ according to our experience $-$ well beyond their reach. We are thus witnessing a shift in M2M connectivity technologies which is being discussed in the remainder of this article.  

\subsection{Low-Power WiFi Technology}

Over the recent years, WiFi (IEEE 802.11) technology has experienced tremendous growth and has become a de-facto solution for home and corporate connectivity. However, WiFi has mostly been out of reach for M2M communications due to its fairly large energy consumption. This has changed as of late, i.e. when the IEEE 802.11 community started to apply duty cycling and hardware optimization with the result of an extremely energy efficient system. 

On the downside, support of mobility and roaming in WiFi is currently rather poor. In terms of reliability, there is neither guaranteed QoS support, nor adequate tools to combat severe interference typical for unlicensed bands. To this end, it has soon been recognized that the favorable propagation properties of low-frequency spectrum at sub-1GHz may provide improved communication properties when compared to, e.g., conventional WiFi protocols operating at 2.4 and 5GHz bands. However, the available spectrum at sub-1GHz license-exempt ISM bands is extremely scarce and hence required careful system design considerations.

With this in mind, after outlining the purpose and the technical scope of the novel IEEE 802.11ah project, the standardization work of the corresponding TGah task group has commenced in November 2010. The prospective technology is generally based on a variation of IEEE 802.11ac standard, but down-clocked by a factor of 10. It is currently being developed to enable low-cost long-range (up to 1km) connectivity across massive M2M deployments with high spectral and energy efficiencies. Today, thousands of M2M devices may already be found in dense urban areas, which required providing support for up to 6K machines connecting to a single access point.

Fortunately, IEEE 802.11ah technology does not need to maintain backward compatibility with the other representatives of the IEEE 802.11 family. Operating over different frequencies, 802.11ah could thus afford defining novel compact frame formats, as well as offering more efficient mechanisms to support a large population of devices, advanced channel access schemes, as well as important power saving and throughput enhancements~\cite{11ah}. As the result, 802.11ah is believed to significantly enrich the family of 802.11 protocols, which already receive increasing attention from mobile network operators willing to introduce low-cost connectivity in unlicensed bands.

\subsection{Unlicensed Low-Power Wide Area Networks}
Given that Zigbee-like solutions have not lived up to their expectations, whereas Low-Power WiFi and Cellular M2M systems have commenced to take shape only recently, a novel class of M2M technologies has emerged lately, termed Low-Power Wide Area (LPWA), which operate in unlicensed spectrum. Only low data rates and small daily traffic volumes are however foreseen~\cite{ETSI2014}, which limits the application to a subset of M2M services with infrequent small data transmissions. 

LPWA technology today is proprietary with multiple non-compatible alternatives. There are also initiatives to propose LPWA technology concepts into the cellular M2M direction within a recent new study item that has been initiated in the 3GPP GERAN (GSM/EDGE Radio Access Network) specification group. Similar to the standardization targets of LTE evolution in 3GPP Radio Access Network (RAN), the motivation is to enable extended coverage beyond GSM coverage today, low device complexity, and long battery lifetimes.

Our experience with LPWA shows that it works successfully for large projects, such as the Moscow Smart City deployment~\cite{Moscow}, where almost 20K sensors have been connected to a modest number of access points. In the trial deployments, we have seen suburban and rural ranges of over 20km, the typical urban ranges of around 5km, and the ``difficult'' urban ranges of 1-2km. Mobile network operators may hence become the early adopters of this emerging technology building on their well-developed network infrastructure and strong customer trust. For instance, a possible deployment model for an operator may be to install LPWA systems complementary to existing cellular technology and the cell sites that they already have~\cite{Mac14}.  

Despite the time-to-market benefits of LPWA, there are also clear downsides of using unlicensed spectrum for long-range communication. Typical regulation imposes several restrictions on radio transmitters in unlicensed spectrum~\cite{ETSI12} in terms of effective radiated power (ERP), allowed duty cycles, and listen-before-talk requirements. For long-range transmissions, the limited ERP causes asymmetric link budgets between uplink and downlink directions. The reason is that the ERP is limiting the radiated power after the antenna gain has been applied. However, antennas have significantly different performance between simple devices using a single antenna with around 0dB antenna gain and a base station with an antenna gain of around 19dB. This means that the uplink signal has  an additional antenna gain at the receiver in contrast to the downlink signal. For European regulation, this can be partly compensated by selecting a subband for downlink with 13dB higher output power. But even then a link asymmetry of at least 6dB remains.

As a consequence, at least 50\% of devices experience only uplink connectivity under non-line-of-sight propagation conditions. This is unreliable in the sense that no acknowledgements for successful uplink data delivery are possible. Further, scalability limitations come from the range covered by a single LPWA base station~\cite{ETSI2014}. Projecting that the total number of connected M2M devices is to become approximately 10 times larger than the number of people, easily millions of devices may appear within the coverage area of a single LPWA base station. Many of those will use other radio technologies that share the spectrum with LPWA, such as low-power WiFi (IEEE 802.11ah), Z-Wave, Zigbee, IEEE 802.15.4g, etc. With its low receiver sensitivity for long-range communication, the LPWA device will perceive all of these other transmissions as interference. 

We therefore foresee that LPWA will only remain viable at the early stage of IoT development when the number of devices is still moderate. However, LPWA can play an important role to support the early IoT market up-take until standardized cellular M2M solutions enter the market, which can handle the anticipated IoT scale $-$ in terms of numbers of devices, but also the variety of M2M services.

\subsection{Cellular M2M}

Cellular technologies, and especially 3GPP LTE, are becoming increasingly attractive for supporting large-scale M2M installations due to their wide coverage, relatively low deployment costs, high level of security, access to dedicated spectrum, and simplicity of management. However, LTE networks have been neither historically designed with link budget requirements of M2M devices, nor optimized for M2M traffic patterns. Therefore, several improvements targeting M2M solutions have been initiated in 3GPP aiming at augmenting LTE to become more suitable for M2M applications. 

Given that the numbers of connected machines are expected to grow dramatically, LTE technology requires respective mechanisms to handle a very large number of devices~\cite{Zhe14}. Correspondingly, an overload control scheme named Enhanced Access Barring has been introduced as part of LTE Rel-11 to avoid overload in RAN, whenever there is a surge in near-simultaneous network entry attempts. Further, accounting for the fact that typical M2M data transmissions are infrequent and small, simplified signaling procedures for radio-bearer establishment are necessary to offer energy consumption savings for such M2M devices. In connection to lightweight signaling for small data, M2M device energy consumption can be reduced significantly for infrequent traffic by allowing for longer cycles of discontinuous reception (DRX).

In the following sections, we review some of these important improvements in more detail. We intentionally focus our description on LTE which we believe will become the major technology for M2M connectivity even though M2M-centric improvements are being discussed for other 3GPP technologies as well.

\begin{figure*}[!ht]
\centering
\includegraphics[width=1.95\columnwidth]{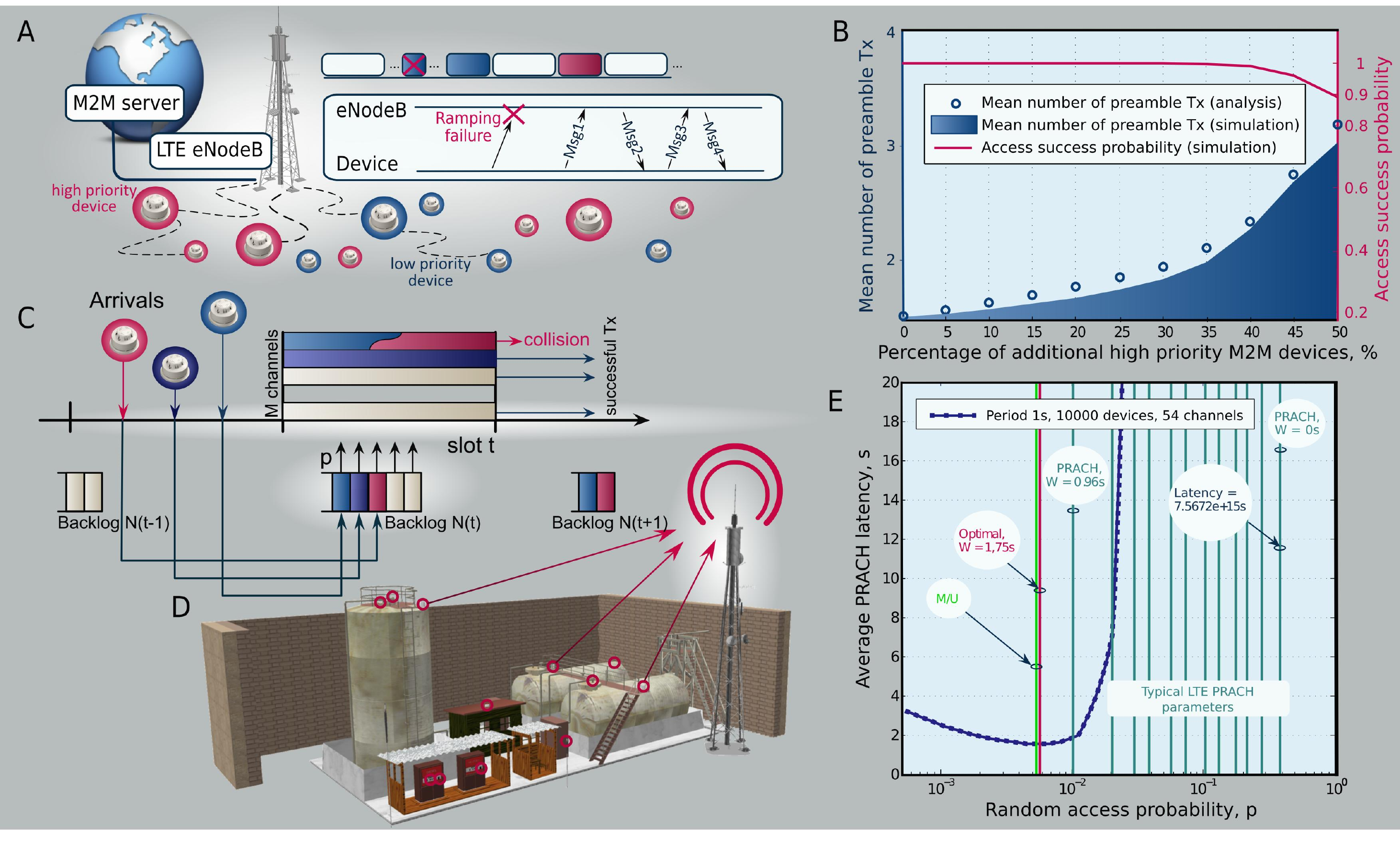}
\caption{Performance results: handling very large numbers of devices. A: motivating M2M scenario; B: connected-mode performance of different device priorities; C: proposed multi-channel M2M contention model; D: characteristic industrial automation application; E: analytical random-access latencies. }
\label{fig:results1}
\end{figure*}

\section{M2M Performance Improvements by 3GPP}

\subsection{Handling Very Large Numbers of Devices}

Our research indicates that smart grid is one of the key M2M use cases incorporating a large number of metering devices that autonomously report their information to grid infrastructure. The motivating smart metering use case therefore serves as a valuable reference ``massive M2M'' scenario covering many characterizing M2M features (see Fig.~\ref{fig:results1}.A). Correspondingly, the involved M2M devices may be divided into several classes according to the priority of their information, e.g. high-priority (alarm messages) and low-priority (measurement data). Potentially, alarm messages constitute a bigger challenge for the network to handle, as they are typically highly synchronized and in addition may require certain latency guarantees~\cite{Con15-2}.

Currently, 3GPP LTE system defines a number of communication channels to deliver uplink transmissions from M2M devices to the network. In particular, the Physical Random Access Channel (PRACH) is employed by a device for its initial network entry, as well as to demand system resources if it does not already have a dedicated resource allocation. In case of many M2M devices connecting to the network near-simultaneously, we expect that the use of PRACH would be preferred, but may result in congestion due to its insufficient capacity. 

More specifically, the PRACH procedure features two distinct stages (see Fig.~\ref{fig:results1}.A). The former is the uplink timing synchronization stage (known as Msg1/Msg2), where the power ramping technique may be used to adjust the transmit power of a random-access preamble to particular channel conditions. Further, Msg3 is used to transmit a meaningful uplink message to the base station (termed eNodeB or eNB) and Msg4 is utilized for subsequent contention resolution.

To understand the impact of a large population of M2M devices on their network access, we construct an event-driven protocol-level PRACH simulator and thoroughly calibrate it against the reference 3GPP methodology documents~\cite{TR11}. Our simulation yields important conclusions on overloaded PRACH performance, when numerous \textit{connected-mode} M2M devices of different priorities send their information into the network (see Fig.~\ref{fig:results1}.B). In particular, we learn that around 40\% of high-priority M2M devices, added to the original (typical) population of 30K low-priority devices, produce a sharp degradation in network access success probability. 

Interestingly, PRACH preambles selected by the M2M devices randomly may be regarded as non-interfering code-based ``channels'' (see Fig.~\ref{fig:results1}.C), where the case when two or more devices select an identical preamble (channel) would correspond to a conventional ``packet'' collision. This opens door to assessing the contention-based M2M behavior by relying on past knowledge of multichannel random-access protocols. 

First, careful custom-made approximations can be forged for particular given ranges of PRACH parameters (see Fig.~\ref{fig:results1}.B), such as the number of available preambles ($M$) and contending devices ($U$), backoff window size, etc. However, these may not be counted as adequate universal solutions and another alternative is straightforward numerical analysis of contention behavior, which would only remain feasible for moderate numbers of users/channels due to high computational complexity.

More recently, we have demonstrated the feasibility of applying powerful fluid approximation techniques to rigorously characterize M2M performance, as well as the stability regions of a multichannel random-access system. As our target scenario, we have chosen an industrial automation application (see Fig.~\ref{fig:results1}.D), which may require certain data access latency and reliability guarantees (e.g., for supporting priority or critical alarm messages). Along these lines, Fig.~\ref{fig:results1}.E indicates \textit{analytical} PRACH latencies as evaluated with our method, which allows optimizing channel access by properly selecting the Msg1 retransmission probability value for arbitrary numbers of devices and channels. 

More specifically, in the figure we compare our optimized latency against the values produced with the use of existing PRACH backoff indicator parameters. Our solution thus helps the base station regulate PRACH access by having system-wide knowledge across all connected M2M devices. However, if such knowledge is not available, simpler heuristic access control procedures (such as when the retransmission probability is chosen as $M$ over $U$) may be employed by the M2M devices locally, which sometimes results in close to optimal performance. These results allow for tighter control of important performance indicators, such as data access latency, which may benefit LTE in supporting constrained automation scenarios on the way to the Industrial Internet.

\begin{figure*}[!ht]
\centering
\includegraphics[width=1.95\columnwidth]{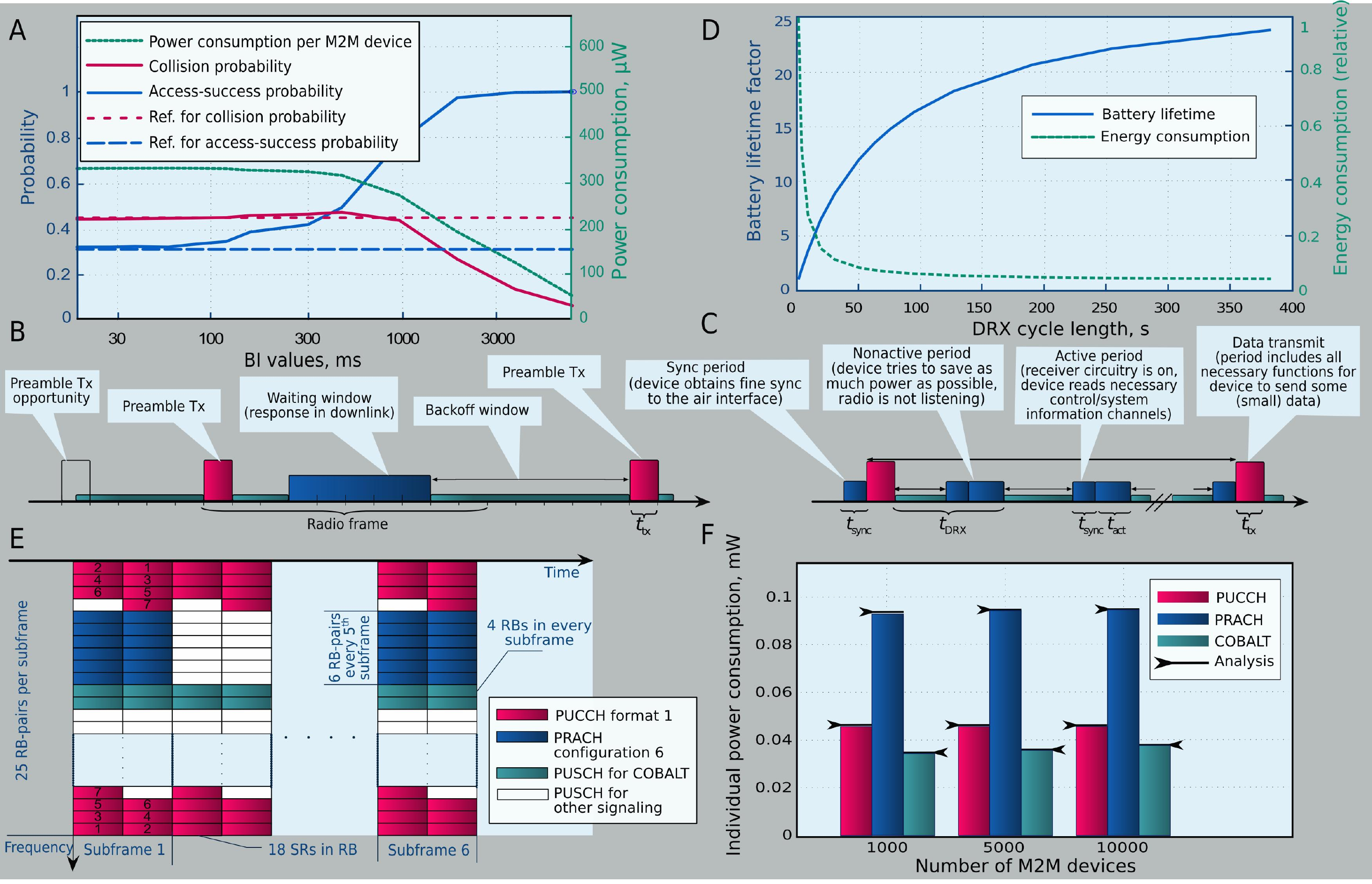}
\caption{Performance results: energy efficiency and small data. A: initial network entry performance; B: M2M device power consumption levels; C: proposed power consumption model; D: assessment of energy efficiency M2M improvements; E: basic LTE frame structure; F: benefits of COBALT mechanism.}
\label{fig:results2}
\end{figure*}

\subsection{Energy Efficiency and Small Data Transmission}

In tight connection with access latency and success rate of M2M transmissions goes their energy efficiency, which is accentuated by the fact that M2M devices are typically small-scale and battery-powered. We continue studying the scenario when an IoT application requires a large number of M2M devices to perform a particular action near-simultaneously (e.g., smart meter data readings), or when an unexpected surge, outage, or failure occurs (massive power outage or restoration of power, network failure, etc.) causing multiple devices to (re)connect to the network within a short period of time. In this case, the transmitting devices would still be using the PRACH contention-based random access procedure to obtain uplink synchronization for initial network access or respective data transmission.

Along these lines, Fig.~\ref{fig:results2}.A illustrates the simulated initial network entry performance of 30K M2M devices with respect to their power consumption, collision probability, and access success probability across different PRACH backoff indicator (BI) values. These results are produced for characteristic beta-distributed M2M device activation patterns (traffic type 2: beta distribution over 10 seconds), as suggested by 3GPP evaluation methodology (see Table 6.1.1 in~\cite{TR11}), since uniformly-distributed activations (traffic type 1: uniform distribution over 60 seconds) do not cause actual network overloads. Our evaluation framework accounts for the main M2M device power consumption levels (inactive, idle, Rx, and Tx) at all states of PRACH signaling procedure (see Fig.~\ref{fig:results2}.B) and sheds light on the feasibility of candidate network overload control solutions. In particular, Fig.~\ref{fig:results2}.A suggests that a combination of M2M-specific backoff (larger non-standard BI values) and initial backoff (pre-backoff) may successfully alleviate congestion caused by highly-correlated beta-distributed M2M device activation patterns.

Further, the focus of our investigation shifts to the dedicated power consumption aspects of M2M devices. Currently, short paging cycles in 3GPP LTE may be highly sub-optimal for M2M devices, especially given lengthy M2M traffic inter-arrival times and delay tolerant nature of many M2M applications. Hence, extending paging cycle durations in the idle state may help delay-tolerant devices sleep for longer periods of time thus extending their battery lifetimes. The corresponding studies require an appropriate power consumption model (see Fig.~\ref{fig:results2}.C), which is capable of capturing typical M2M traffic patterns. Correspondingly, our results in~\cite{Tir13} indicate that increasing the current maximum DRX (discontinuous reception) and paging cycle lengths would indeed lead to significant gains in the energy consumption (over 20x) of M2M devices (see Fig.~\ref{fig:results2}.D). 

If additional delay can be tolerated by M2M devices, D2D (device-to-device) communication techniques may further reduce the consumption of power. For instance, one M2M device may act as an aggregation point and relay data from other proximate M2M devices with poor communication link (to avoid excessive retransmissions and associated energy costs). D2D-based ``client relay'' mechanisms may dramatically reduce energy expenditures of cell-edge M2M devices, especially when those only send small data packets, and additionally help relieve a surge in uncontrolled near-simultaneous M2M transmissions. However, if adding more delay is not acceptable, such as in critical control applications, further data access improvements would need to be made reducing PRACH latencies to few milliseconds by, potentially, shortening the signaling sequence in Fig.~\ref{fig:results2}.B. Other areas for data access enhancements concern the size of typical M2M payloads (on the order of several bytes), as existing coding mechanisms in LTE are not optimal for short data blocks. Small data also creates inefficiency in control and channel estimation procedures causing excessive overheads, as well as in existing frame structures and resource allocation schemes.

In more detail, Fig.~\ref{fig:results2}.E illustrates the current frame structure of LTE (for 5MHz bandwidth) as a rectangular grid of resource blocks (RBs). The groups of RBs compose different data access channels, including periodic PRACH allocations and continuous Physical Uplink Control Channel (PUCCH) resources to carry the uplink control information. As both PUCCH and PRACH capacities may be very limited to serve small M2M data from numerous sources, we propose to allocate a part of Physical Uplink Shared Channel (PUSCH), which is otherwise employed for actual human-to-human (H2H) data transmissions, for a dedicated M2M use. Our respective scheme, named contention-based LTE transmission (COBALT), takes advantage of fewer LTE signaling messages and a simple collision resolution procedure (see detailed description in~\cite{And13}). It thus yields better utilization of network resources, lower latencies, and, most importantly, significantly reduced power consumption for M2M devices (see Fig.~\ref{fig:results2}.F).


\section{3GPP Standards Update and Future Outlook}
We conclude this work by offering the current standardization perspective on cellular M2M technologies. Even though presently there is a number of industrial alliances and research projects pursuing their own standards and technologies in the M2M space, in this work we concentrate primarily on cellular sector due to the major promise that it holds for the entire IoT industry.

M2M or machine-type communication was identified as one of the major topics for further enhancement in 3GPP due to an extreme diversity of prospective applications and corresponding consumer demands. In particular, the RAN technical specification group of 3GPP has been very active on M2M-related features across several releases of LTE as well as legacy cellular technologies. To this end, LTE Rel-11 has introduced improvements in handling a large number of M2M devices with delay-tolerant traffic, including RAN overload control mechanisms (such as Enhanced Access Barring) and device power preference indication. In LTE Rel-12, 3GPP had a number of M2M-related study items (e.g., UEPCOP on power saving optimizations and SDDTE on signaling enhancements). The most significant outcomes of this work are a dedicated power saving mode (providing possibility for battery lifetime savings for M2M, potentially over 10 years of operation) and a new low-complexity device category, named Cat-0.

As the result of past 3GPP work, the M2M devices can be flagged as low-priority and barred from accessing the network in case it is congested. In addition, scheduling prioritization and service differentiation mechanisms have been ratified to efficiently handle different traffic types by minimizing the impact of M2M data on H2H traffic. Further, low-complexity device modules support LTE operation with a single receiver chain and antenna, reduced peak data rates of 1Mbps, and optional half-duplex operation. This work is planned to continue in Rel-13 along the lines of further coverage and power saving enhancements, as well as a new device category based on Cat-0, but with even lower complexity (reduced RF bandwidth of 1.4MHz and maximum transmit power of 20dBm compared to 23dBm of today). Small data transmission and coverage enhancements are considered by some study/work items, together with architectural latency-related modifications.

Consequently, cellular networks are becoming increasingly equipped to support a diversity of M2M use cases and associated technical requirements $-$ they already offer sufficient bandwidths and nearly-ubiquitous coverage (further improved by 15-20dB in Rel-12/13), support full mobility and provide precise location information. With the ongoing LTE evolution and the corresponding thorough standardization, (i) M2M traffic can coexist efficiently with H2H mobile broadband applications~\cite{Con15}, (ii) M2M modem complexity drops by 50\% (Rel-12) and by up to 75\% (Rel-13) comparing to today's cheapest Cat-1 UE, thus resulting in lower modem costs, and (iii) battery lifetimes extend over 10 years for downlink delay-tolerant traffic (Rel-12) and other use cases (planned in Rel-13). In turn, lower complexity UE categories (Rel-12 and the upcoming Rel-13 work) provide attractive cost saving opportunities for the chipset manufacturers.

As our world is moving towards a fully-integrated networked society, where everything that may benefit from interacting and sharing information will become connected, 5G-grade M2M systems are expected sometime around 2020. They should generally deliver ubiquitous M2M connectivity with edge-free experience, either on a stand-alone Cellular M2M carrier or multiplexed with other services (e.g., mobile broadband). Given that wireless connectivity is becoming a new commodity, same as water or electricity, M2M-based applications are likely to become a centerpiece of the emerging 5G ecosystem by enabling ubiquitous interworking between various communicating objects, as well as collection and sharing of the massive amounts of data. For the research community, however, these emerging future systems come with their associated unique challenges, such as extreme heterogeneity of services, large-scale unattended wireless connectivity, and unprecedentedly large volumes of information to handle.

\bibliographystyle{ieeetr}
\bibliography{refs}

\begin{thebibliography}{10}

\bibitem{SRA11}
Finnish Strategic Centre for Science, Technology, and Innovation, {\em Internet
  of Things Strategic Research Agenda (IoT-SRA)}, September 2011.

\bibitem{Sto14}
I.~Stojmenovic, ``Machine-to-machine communications with in-network data
  aggregation, processing, and actuation for large-scale cyber-physical
  systems,'' {\em IEEE Internet of Things Journal}, vol.~1, no.~2,
  pp.~122--128, 2014.

\bibitem{Sac14}
J.~Sachs, N.~Beijar, P.~Elmdahl, J.~Melen, F.~Militano, and P.~Salmela,
  ``Capillary networks -- a smart way to get things connected,'' {\em Ericsson
  Review}, vol.~8, pp.~1--8, 2014.

\bibitem{Ast13}
D.~Astely, E.~Dahlman, G.~Fodor, S.~Parkvall, and J.~Sachs, ``{LTE Release 12}
  and beyond,'' {\em IEEE Communications Magazine}, vol.~51, no.~7,
  pp.~154--160, 2013.

\bibitem{Moscow}
I.~Vilajosana, J.~Llosa, B.~Martinez, M.~Domingo-Prieto, A.~Angles, and
  X.~Vilajosana, ``Bootstrapping smart cities through a self-sustainable model
  based on big data flows,'' {\em IEEE Communications Magazine}, vol.~51,
  no.~6, pp.~128--134, 2013.

\bibitem{11ah}
{\em IEEE, "TGah Functional Requirements and Evaluation Methodology"}.

\bibitem{ETSI2014}
{\em ETSI GS LTN 001, "Low Throughput Networks (LTN): Use Cases, Functional
  Architecture and Protocols"}.

\bibitem{Mac14}
Machina Research, {\em The need for low cost, high reach, wide area
  connectivity for the {Internet of Things}}, 2014.

\bibitem{ETSI12}
{\em ETSI EN 300 200-1, "Electromagnetic compatibility and Radio spectrum
  Matters (ERM); Short Range Devices (SRD); Radio equipment to be used in the
  25 MHz to 1 000 MHz frequency range with power levels ranging up to 500 mW;
  Part 1: Technical characteristics and test methods"}.

\bibitem{Zhe14}
K.~Zheng, S.~Ou, J.~Alonso-Zarate, M.~Dohler, F.~Liu, and H.~Zhu, ``Challenges
  of massive access in highly dense {LTE-advanced} networks with
  machine-to-machine communications,'' {\em IEEE Wireless Communications},
  vol.~21, no.~3, pp.~12--18, 2014.

\bibitem{Con15-2}
M.~Condoluci, M.~Dohler, G.~Araniti, A.~Molinaro, and J.~Sachs, ``Enhanced
  radio access and data transmission procedures facilitating industry-compliant
  machine-type communications over {LTE}-based {5G} networks,'' in {\em IEEE
  Wireless Communications}, 2015.

\bibitem{TR11}
{\em Study on RAN improvements for machine-type communications. 3GPP TR 37.868,
  2011}.

\bibitem{Tir13}
T.~Tirronen, A.~Larmo, J.~Sachs, B.~Lindoff, and N.~Wiberg,
  ``Machine-to-machine communication with long-term evolution with reduced
  device energy consumption,'' {\em Transactions on Emerging Telecommunications
  Technologies}, vol.~24, no.~4, pp.~413--426, 2013.

\bibitem{And13}
S.~Andreev, A.~Larmo, M.~Gerasimenko, V.~Petrov, O.~Galinina, T.~Tirronen,
  J.~Torsner, and Y.~Koucheryavy, ``Efficient small data access for
  machine-type communications in {LTE},'' in {\em Proc. of the IEEE
  International Conference on Communications (ICC)}, pp.~3569--3574, 2013.

\bibitem{Con15}
M.~Condoluci, M.~Dohler, G.~Araniti, A.~Molinaro, and K.~Zheng, ``Toward {5G
  DenseNets}: Architectural advances for effective machine-type communications
  over femtocells,'' {\em IEEE Communications Magazine}, vol.~53, no.~1,
  pp.~134--141, 2015.

\end{thebibliography}

{\small
\section*{Authors' Biographies}

\textbf{Sergey Andreev} (sergey.andreev@tut.fi) is a Senior Research Scientist in the Department of Electronics and Communications Engineering at Tampere University of Technology, Finland. He received the Specialist degree (2006) and the Cand.Sc. degree (2009) both from St. Petersburg State University of Aerospace Instrumentation, St. Petersburg, Russia, as well as the Ph.D. degree (2012) from Tampere University of Technology. Sergey (co-)authored more than 90 published research works on wireless communications, energy efficiency, heterogeneous networking, cooperative communications, and machine-to-machine applications.

\textbf{Olga Galinina} (olga.galinina@tut.fi) is a Ph.D. Candidate in the Department of Electronics and Communications Engineering at Tampere University of Technology, Finland. She received her B.Sc. and M.Sc. degrees in Applied Mathematics from Department of Applied Mathematics, Faculty of Mechanics and Physics, St. Petersburg State Polytechnical University, Russia. Her research interests include applied mathematics and statistics, queuing theory and its applications; wireless networking and energy efficient systems, machine-to-machine and device-to-device communication.

\textbf{Alexander Pyattaev} (alexander.pyattaev@tut.fi) is a Ph.D. Candidate in the Department of Electronics and Communications Engineering at Tampere University of Technology, Finland. He received his B.Sc. degree from St. Petersburg State University of Telecommunications, Russia, and his M.Sc. degree from Tampere University of Technology. Alexander has publications on a variety of networking-related topics in internationally recognized venues, as well as several technology patents. His primary research interest lies in the area of future wireless networks: shared spectrum access, smart RAT selection and flexible, adaptive topologies.

\textbf{Mikhail Gerasimenko} (mikhail.gerasimenko@tut.fi) is a Researcher at Tampere University of Technology in the Department of Electronics and Communications Engineering. Mikhail received Specialist degree from Saint-Petersburg University of Telecommunications in 2011. In 2013, he obtained Master of Science degree from Tampere University of Technology. Mikhail started his academic career in 2011 and since appeared as (co-)author on multiple scientific journal and conference publications, as well as several patents. His main subjects of interest are wireless communications, machine-type communications, and heterogeneous networks.

\textbf{Tuomas Tirronen} (tuomas.tirronen@ericsson.com) is a Senior Researcher at Ericsson Research, which he joined in 2012. He received his D.Sc. in Communications Engineering in 2010 from Aalto University. His current research interests include 4G and 5G wireless access technologies, Internet of Things, performance evaluation, radio protocols and resources. He is also active in 3GPP standardization work and innovation and patenting.

\textbf{Johan Torsner} (johan.torsner@ericsson.com) is a Research Manager in Ericsson Research and is currently leading Ericsson's research activities in Finland. He joined Ericsson in 1998 and has held several positions within research and R\&D. He has been deeply involved in the development and standardization of 3G and 4G systems and has filed over 100 patent applications. His current research interests include 4G evolution, 5G and machine-type communication.

\textbf{Joachim Sachs} (joachim.sachs@ericsson.com) is a Principal Researcher at Ericsson Research. He joined Ericsson in 1997 and has worked on a variety of topics in the area of wireless communication systems. He holds a diploma in electrical engineering from Aachen University (RWTH), and a doctorate in electrical engineering from the Technical University of Berlin, Germany. Since 1995 he has been active in the IEEE and the German VDE Information Technology Society (ITG), where he is currently co-chair of the technical committee on communication.

\textbf{Mischa Dohler} (mischa.dohler@kcl.ac.uk) is Chair Professor in Wireless Communications at King's College London, Director of the Centre for Telecommunications Research, co-founder and member of the Board of Directors of the smart city pioneer Worldsensing, Fellow (2014) and Distinguished Lecturer of the IEEE, and Editor-in-Chief of the Transactions on Emerging Telecommunications Technologies. He is a frequent keynote, panel and tutorial speaker. He has pioneered several research fields, contributed to numerous wireless broadband, IoT/M2M and cyber security standards, holds a dozen patents, organized and chaired numerous conferences, has more than 200 publications, and authored several books. He has a citation h-index of 39 (top 1\%). He acts as policy, technology and entrepreneurship adviser, examples being Richard Branson's Carbon War Room, House of Lords UK, UK Ministry BIS, EPSRC ICT Strategy Advisory Team, European Commission, ISO Smart City working group, and various start-ups. He is also an entrepreneur, angel investor, passionate pianist and fluent in 6 languages. He has talked at TEDx. He had coverage by national and international TV \& radio; and his contributions have featured on BBC News and the Wall Street Journal.

\textbf{Yevgeni Koucheryavy} (yk@cs.tut.fi) is a Full Professor and Lab Director at the Department of Electronics and Communications Engineering of Tampere University of Technology (TUT), Finland. He received his Ph.D. degree (2004) from TUT. He is the author of numerous publications in the field of advanced wired and wireless networking and communications. His current research interests include various aspects in heterogeneous wireless communication networks and systems, the Internet of Things and its standardization, as well as nanocommunications. He is Associate Technical Editor of IEEE Communications Magazine and Editor of IEEE Communications Surveys and Tutorials.

}

\end{document}